\documentclass[sigconf]{acmart}

\usepackage{multirow} 
\usepackage{makecell} 
\usepackage{subcaption}
\usepackage{tabularx} 

\AtBeginDocument{%
  }

\copyrightyear{2024}
\acmYear{2024}
\setcopyright{acmlicensed}
\acmConference[CIKM '24] {Proceedings of the 33rd ACM International Conference on Information and Knowledge Management}{October 21--25, 2024}{Boise, ID, USA.}
\acmBooktitle{Proceedings of the 33rd ACM International Conference on Information and Knowledge Management (CIKM '24), October 21--25, 2024, Boise, ID, USA}
\acmISBN{979-8-4007-0436-9/24/10}
\acmDOI{10.1145/3627673.3680055}

\acmSubmissionID{1309}

\begin{document}

\title{EXIT: An EXplicit Interest Transfer Framework for Cross-Domain Recommendation}

\author{Lei Huang}
\affiliation{
  \institution{Meituan}
  \city{Beijing}
  \country{China}
  }
\email{huanglei45@meituan.com}

\author{Weitao Li}
\affiliation{
  \institution{Meituan}
  \city{Beijing}
  \country{China}
  }
\email{liweitao05@meituan.com}

\author{Chenrui Zhang}
\affiliation{
  \institution{Meituan}
  \city{Beijing}
  \country{China}
  }
\email{chenrui.zhang@pku.edu.cn}

\author{Jinpeng Wang}
\affiliation{
  \institution{Meituan}
  \city{Beijing}
  \country{China}
  }
\email{wangjinpeng04@meituan.com}

\author{Xianchun Yi}
\affiliation{
  \institution{Meituan}
  \city{Beijing}
  \country{China}
  }
\email{yixianchun@meituan.com}

\author{Sheng Chen}
\affiliation{
  \institution{Meituan}
  \city{Beijing}
  \country{China}
  }
\email{chensheng19@meituan.com}

\renewcommand{\shortauthors}{Lei Huang et al.}

\begin{abstract}
Cross-domain recommendation has attracted substantial interest in industrial apps such as Meituan, which serves multiple business domains via knowledge transfer and meets the diverse interests of users. However, existing methods typically follow an implicit modeling paradigm that blends the knowledge from both the source and target domains, and design intricate network structures to share learned embeddings or patterns between domains to improve recommendation accuracy. Since the transfer of interest signals is unsupervised, these implicit paradigms often struggle with the negative transfer resulting from differences in service functions and presentation forms across different domains. In this paper, we propose a simple and effective \textbf{EX}plicit \textbf{I}nterest \textbf{T}ransfer framework named \textbf{EXIT} to address the stated challenge. Specifically, we propose a novel label combination approach that enables the model to directly learn beneficial source domain interests through supervised learning, while excluding inappropriate interest signals. Moreover, we introduce a scene selector network to model the interest transfer intensity under fine-grained scenes. Offline experiments conducted on the industrial production dataset and online A/B tests validate the superiority and effectiveness of our proposed framework. Without complex network structures or training processes, EXIT can be easily deployed in the industrial recommendation system. EXIT has been successfully deployed in the online homepage recommendation system of Meituan App, serving the main traffic.
\end{abstract}




\maketitle

\begin{figure}[ht]
\vspace{-0.4cm}
\setlength{\abovecaptionskip}{-1pt} 
\setlength{\belowcaptionskip}{-1pt}
\centering
\includegraphics[width=0.45\textwidth]{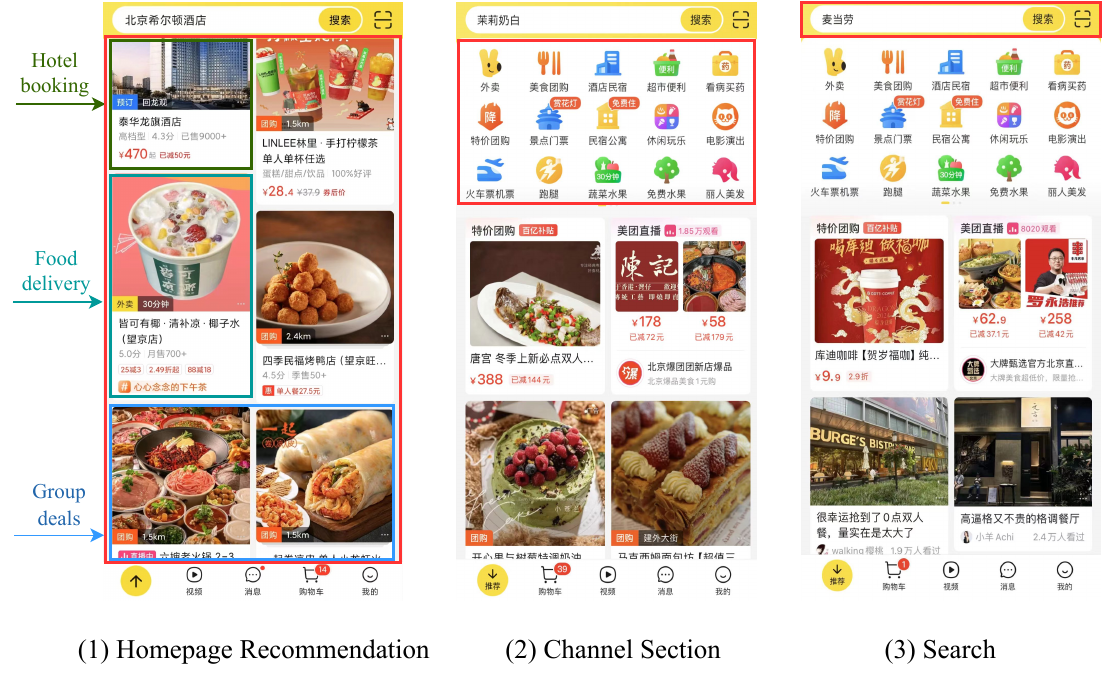} 
\vspace{0.1cm}
\caption{Three representative business domains on Meituan mobile app: \textit{Homepage Recommendation}, \textit{Channel Section}, and \textit{Search}. Each domain offers a variety of businesses, including hotel booking, food delivery, group deals, etc.}
\label{fig:domain reprensent}
\vspace{-0.5cm}
\end{figure}

\section{Introduction}

Recommendation systems have received long-term and widespread attention in both academia and industry. Traditional recommendation methods~\cite{zhou2018deep, ma2018entire, GuoTYLH17, xie2021deep, WangCLHCYLC22} often focus on a single domain, using data from a single domain to mine user interests. However, due to the sparsity of most user behaviors and incomplete user interests in a single domain, single-domain recommendation methods suffer from the long-standing data sparsity problem~\cite{natarajan2020, idrissi2020} and biased interest estimation~\cite{Ouyang20, sun2023remit}. Cross-Domain Recommendation (CDR) methods transfer knowledge from the source domain to the target domain to solve the problems existing in single-domain recommendation, which currently attract continuous exploration. For example, on large e-commerce platforms like Amazon and life service platforms like Meituan, there are multiple business domains to meet the diverse interests of users. Figure~\ref{fig:domain reprensent} illustrates three representative business domains on the Meituan mobile app. CDR methods can be naturally applied to these platforms to improve recommendation accuracy.

However, there are two key challenges when implementing CDR methods in Meituan. Firstly, there is a marked contrast between the service functions of the search domain, which addresses the user's explicit needs, and the recommendation domain, which predicts the user's latent needs. For instance, user's sporadic medicine orders in the search domain when ill, if misinterpreted as regular interest and transferred to the recommendation domain, could result in numerous medicines being recommended when the user is healthy. Such a negative transfer situation could greatly damage the user experience, potentially leading to user attrition or even negative publicity. Secondly, Meituan's diverse range of businesses, including food delivery, in-store services, hotel booking, and healthcare, amplifies the challenge in preventing negative transfer. As users' interests in each business may vary depending on specific contexts (e.g., location and time),  it is crucial to differentiate the significance of these interests across various contexts when transferring them to the recommendation domain. For instance, user interest in food delivery typically spikes during weekday meal times at the office, while interest in in-store entertainment services increases on weekend afternoons in commercial districts. Therefore, the CDR method must be capable of filtering out beneficial source domain signals that are appropriate for the recommendation domain. 

Early CDR methods mainly utilize the similar content information to link different domains for data augmentation~\cite{winoto2008if, tan2014cross}. Recent CDR methods employ machine learning techniques to first learn user/item embeddings~\cite{Ouyang20, ZhaoLXDS20, DLiuLLP20} or rating patterns~\cite{YuanYB19, LoniSLH14} from the source domain and then transfer them to the target domain. Despite promising progress, existing CDR methods are ineffective when applied to scenarios that have significant differences across domains and businesses, such as Meituan. Firstly, existing CDR methods typically facilitate knowledge transfer between similar domains (such as multiple recommendation domains under different scenarios~\cite{ZhuC0LZ19, ShengZ21, YuanYB19, Pan20}) where the impact of negative transfer is minimal. However, they may struggle in scenarios where there is a marked contrast between the source and target domains. Secondly, existing CDR methods primarily focus on the implicit modeling paradigm, as the learning objectives of these methods are to fit the ground truth of the target or source domain interest, without directly supervising the cross-domain transfer of interest signals. As a result, the transfer of interests from the source domain to the target domain is implicit and uncontrollable. This implicit modeling of interest transfer makes it challenging to filter and differentiate interests in the source domain, so it is highly probable that inappropriate or interfering interests will be transferred to the target domain, leading to negative transfer. 

\begin{figure}[t]
\setlength{\abovecaptionskip}{-1pt} 
\setlength{\belowcaptionskip}{-1pt}
\centering
\includegraphics[width=0.47\textwidth]{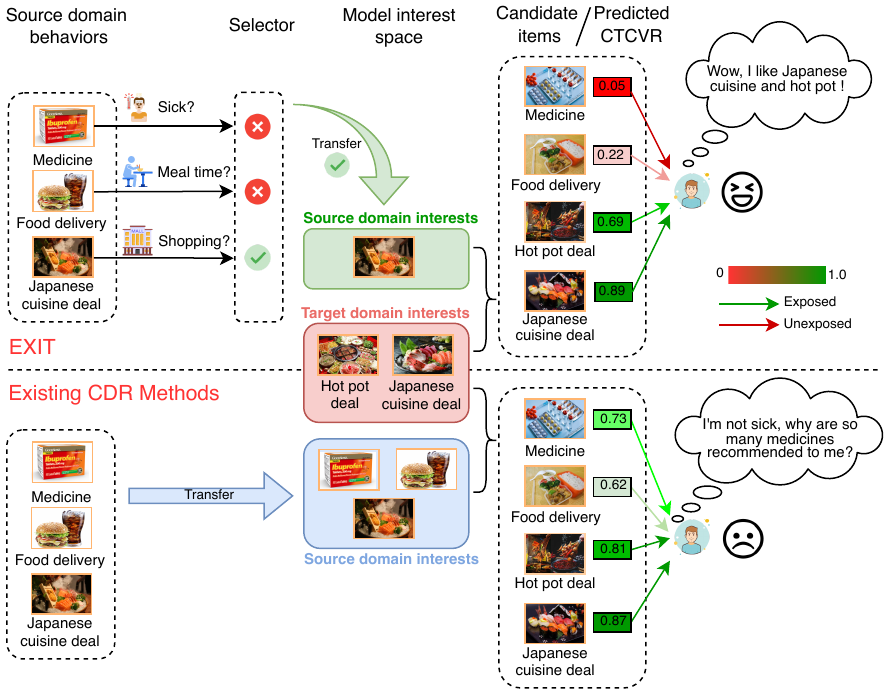} 
\vspace{0.2cm}
\caption{High-level overview of the EXIT. Unlike traditional CDR methods, EXIT explicitly transfers interests from the source domain that are beneficial to the target domain based on user's specific contexts, preventing negative transfer.}
\label{fig:main_idea}
\vspace{-0.4cm}
\end{figure}

To tackle the aforementioned challenges, we propose an \textbf{EX}plicit \textbf{I}nterest \textbf{T}ransfer framework (EXIT) for cross-domain recommendation. As shown in Figure~\ref{fig:main_idea}, EXIT is capable of selectively transferring interests from the source domain based on user's specific contexts (i.e. the fine-grained scenes), ensuring only those beneficial to the target domain are transferred. This enables EXIT to both utilize source domain knowledge to more accurately capture user interests and prevent negative transfer.

The main contributions of this work are summarized as follows:

\begin{itemize}
\item To the best of our knowledge, we are the first to use supervised learning to model the cross-domain interest transfer process, which provides a novel explicit paradigm for cross-domain recommendation.
\item We implement an effective instance for the explicit paradigm. Specifically, we elaborately design the interest combination label, a novel approach for constructing labels for supervised learning of the interest transfer process. To collaborate with the interest combination label, we further propose a scene selector network to model the intensity of cross-domain interest transfer in fine-grained scenes.
\item We conduct extensive experiments on the industrial production dataset and online A/B tests. The consistent superiority validates the effectiveness of EXIT, which is now deployed in the homepage recommendation system of Meituan App. The deployment of EXIT brings 1.23\% CTCVR (Click-Through Conversion Rate) and 3.65\% GTV (Gross Transaction Value) lift, contributing significant improvement to the business and greatly improving the user experience. 
\end{itemize}

\section{Related Work}

\textbf{Single-domain recommendation.} Single-domain recommendation involves training the recommender in a specific business domain and then using it for that domain. To enhance feature extraction, recent recommendation models are moving toward the direction of deep network and feature interaction. DNN~\cite{he2017neural} and PNN~\cite{QuCRZYWW16} are proposed to learn high-order feature representations. Wide\&Deep~\cite{Cheng16} and DeepFM~\cite{GuoTYLH17} integrate shallow techniques and deep network approaches to simultaneously extract both low-order and high-order features. DIN~\cite{zhou2018deep}, DIEN~\cite{ZhouMFPBZZG19}, MINN~\cite{PiBZZG19} and SIM~\cite{PiZZWRFZG20} model the user's historical behavior sequence to capture the diversity of user interests. DCN~\cite{WangFFW17} 
 and DCV-V2~\cite{wang2021dcn} combine the cross network and deep network to enhance feature crossing. Despite their rapid development, single-domain recommendation methods still suffer from the long-standing problems of data sparsity~\cite{natarajan2020, idrissi2020} and biased interest estimation~\cite{Ouyang20, sun2023remit}.

\textbf{Cross-domain recommendation.} CDR methods transfer source domain knowledge to the target domain, alleviating the data sparsity problem and improving recommendation performance in the target domain. Early content-based transfer methods tended to utilize data from source domains to enhance or augment the target domain data~\cite{winoto2008if, tan2014cross, berkovsky2007}. Rating pattern-based transfer methods such as MINDTL~\cite{HeZYY18} and DARec~\cite{YuanYB19} first learn the specific rating patterns of users from the source domain and then transfer the rating patterns to the target domain. Most CDR methods follow an embedding-based transfer paradigm that employs machine learning techniques to learn user/item embeddings and transfer these embeddings across domains. DTCDR~\cite{ZhuC0LZ19} designs an effective embedding-sharing strategy to combine and share the embeddings of common users across domains.  MiNet~\cite{Ouyang20} employs item-level attention and interest-level attention to jointly model user interests from multiple domains. STAR~\cite{ShengZ21} proposes a centered network shared by all domains to model domain commonalities and domain-specific networks tailored for each domain to capture domain characteristics. UniCDR~\cite{cao2023towards} provides a unified framework that solves the data sparsity problem and cold-start problem simultaneously. All of these CDR methods use an unsupervised approach to model cross-domain knowledge transfer.

\section{The Proposed Framework}

\subsection{Problem Setup}
In this paper, we focus on improving CTCVR prediction in recommendation systems by utilizing multi-domain purchase behavior data of users. The CTCVR prediction task in online recommendation systems involves constructing a prediction model to estimate the probability of a user purchasing an item recommended by the system. The predicted CTCVR value \begin{math} \hat{y} \end{math} of a user \begin{math} u \end{math} purchasing an item \begin{math} m \end{math} can be formulated as:
\begin{equation}
   \hat{y} = f \left([E(u_1),.., E(u_i);E(m_1),.. ,E(m_j);E(c_1),.. ,E(c_k)]; \Theta_{f} \right)
    \label{eq:problem setup}
\end{equation}
where \begin{math} f \end{math} is the model function with learnable model parameters \begin{math} \Theta_{f} \end{math}, \begin{math} E(\cdot) \end{math} denotes the embedding layer which maps the raw features into dense vectors. \begin{math} u_i, m_j, c_k \end{math} respectively represent user feature, item feature, and context feature, while \begin{math} i, j \end{math} and \begin{math} k \end{math} denote feature indexes.

The cross-domain CTCVR prediction problem is defined as leveraging the data from source domain to better learn the model function \begin{math} f \end{math} and improve the prediction performance in the target domain. In the industry scenario of this work, the source domain and the target domain share the same set of users and items, as they are different business domains on the Meituan App.

\begin{figure}[t]
  \centering
  \includegraphics[width=0.45\textwidth]{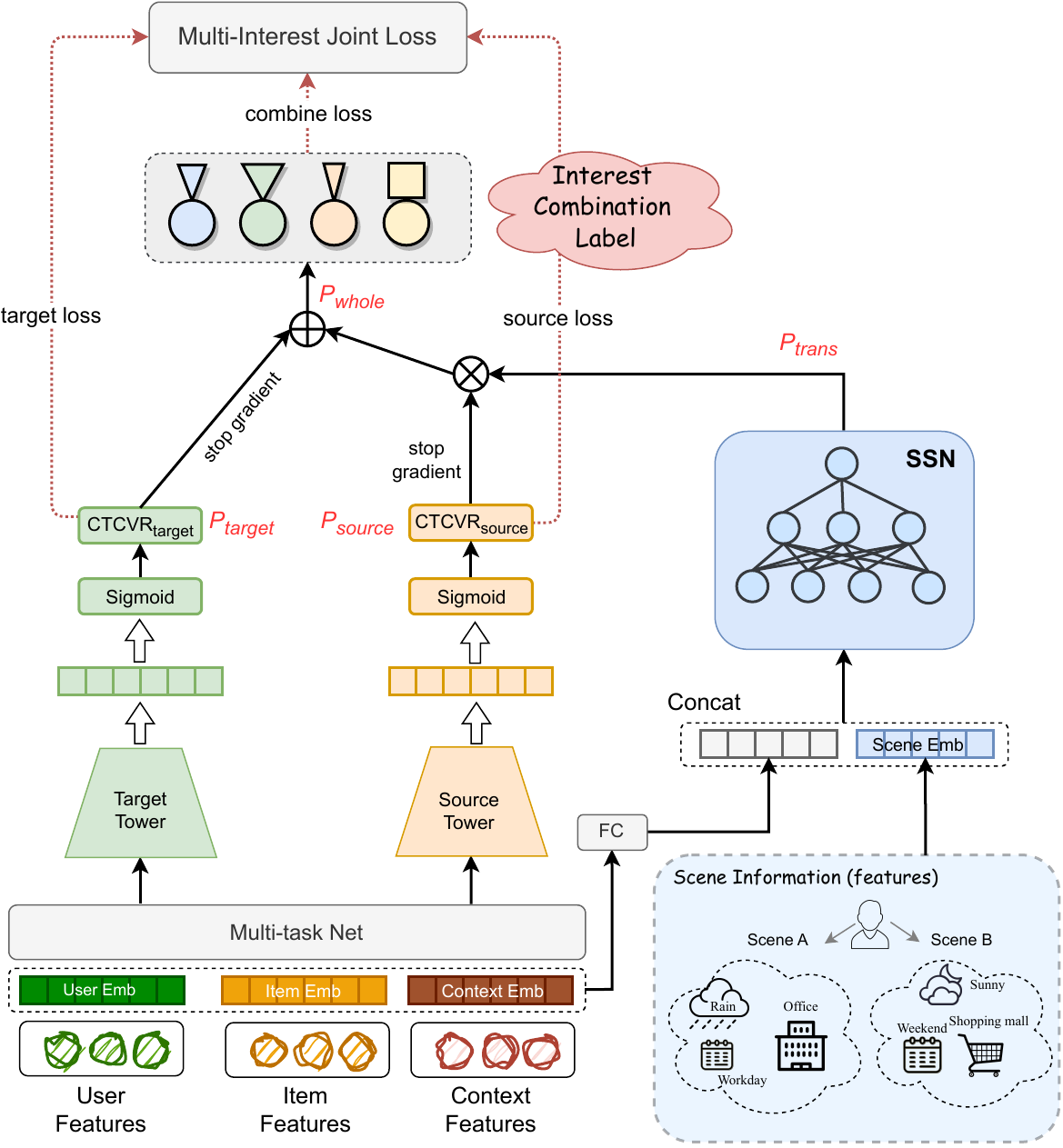}
  \caption{The framework of the EXIT. EXIT models the user's target domain interest \begin{math} P_{target} \end{math}, the source domain interest \begin{math} P_{source} \end{math}, and the cross-domain interest transfer probability \begin{math} P_{trans} \end{math} , ultimately obtaining the user’s complete interest suitable for the target domain as \begin{math} P_{whole} = P_{target}+P_{source}*P_{trans} \end{math}.}
  \Description{Framework}
  \label{fig:framework}
  \vspace{-0.4cm}
\end{figure}

\subsection{The Overall Framework}
The overall framework of EXIT is shown in Figure \ref{fig:framework}. The overall idea is to first independently predict the user's source domain interest and target domain interest, then identify the portion of interests in the source domain that can be appropriately transferred to the target domain through explicit labels, and combine it with the target domain interest to represent the user's complete interest. EXIT is composed of three components: the Interest Prediction Network (IPN), the Interest Combination Label (ICL), and the Scene Selector Network (SSN). IPN models the user's interests in the target and source domains, as the basis for building a comprehensive representation of the user's interests. The role of ICL is to serve as the ground truth of the user's complete interests suitable for the target domain, and as the label for supervised learning in the interest transfer process. SSN helps the model learn in which scenarios it is suitable to carry out cross-domain interest transfer, which takes the scene feature embedding \begin{math} \mathbf{E}^{scene} \end{math} as input and outputs the interest transfer probability \begin{math} P_{trans} \end{math}. These three components work together to fit the user's complete interests suitable for the target domain and eliminate inappropriate interests. From a global perspective, the user's complete interest \begin{math} P_{whole} \end{math} is represented as the interest in the target domain plus the beneficial interest transferred from the source domain, which is formulated as:
\begin{equation}
   P_{whole} = P_{target}+P_{source}*P_{trans}
    \label{eq: global perspective}
\end{equation}
where \begin{math} P_{target}, P_{source} \end{math} respectively represent the user's target domain interest and source domain interest predicted by IPN.

\subsection{Interest Prediction Network}

 As the overlapping users and items between domains share the same basic features, so we model the user interests of different domains based on the multi-task framework. 

\subsubsection{Feature Embedding}

IPN takes user features (user profile, behavior sequence, etc.), item features (item attributes), and context features (hour, weekday, etc.) as inputs. The raw sparse features are mapped into dense embeddings through the embedding layer. For simplicity, user embedding, item embedding, and context embedding are represented as \begin{math} \mathbf{E}^U, \mathbf{E}^M\end{math} and \begin{math} \mathbf{E}^C\end{math}. The feature embeddings are finally transformed into a vector representation \begin{math} \mathbf{V} \end{math} as:
\begin{equation}
   \mathbf{V} = [\mathbf{E}^U || \mathbf{E}^M || \mathbf{E}^C]
\end{equation}
where \begin{math} || \end{math} denotes the vector concatenation operation.

\subsubsection{Multi-domain Interest Aggregation}
We aggregate interests from multiple source domains for a unified representation. Given the original  \begin{math} n \end{math} source domain labels: \begin{math} y_1^s,y_2^s,...,y_n^s \end{math}, where \begin{math} y_i^s \end{math} denotes whether the user has purchased the item in the \begin{math} i \end{math}-th source domain, the aggregate label is represented as:
\begin{equation}
   y^{s} = Max(y_1^s,y_2^s, ... ,y_n^s)
\end{equation}
The aggregation label represents the union interests of multiple source domains. When any source domain label is 1, the aggregate label is 1. As industrial apps typically have dozens of business domains, designing separate networks to consider each source domain would cause the model parameters to inflate, making it difficult to meet the online system's requirements for model memory and inference speed. Unified modeling of source domain interests allows us to only add a source domain interest prediction target in the multi-task network. When the domain expands, the model structure need not be altered. Only the sample labels need to be adapted, facilitating rapid implementation and deployment of the model.

\subsubsection{Interest Prediction Tower}
Taking the feature vector representation \begin{math} \mathbf{V} \end{math} as input, and utilizing the target domain label \begin{math} y^t \end{math} and the source domain aggregation label \begin{math} y^{s} \end{math} as supervision signals, which respectively indicate whether the user has purchased the item in the corresponding domain, the target domain interest tower and the source domain interest tower model the user's target domain interest \begin{math} P_{target} \end{math} and source domain interest \begin{math} P_{source} \end{math} respectively.

\subsection{Interest Combination Label}
Since we use supervised learning to model the cross-domain interest transfer process which requires explicit labels, we construct the ICL as a supervisory signal for whether the source domain signal can be transferred to the target domain. In our study, we use the interest transfer probability \begin{math} P_{trans} \end{math} to quantify whether the signal from the source domain can be transferred to the target domain. During the model training process, by minimizing the loss associated with the ICL, \begin{math} P_{trans} \end{math} achieves the effect of "\textit{tending to 1 when the source domain interest is suitable for the target domain and tending to 0 when it is not}", thus achieving the filtering of the source domain signals, and only transferring beneficial signals to the target domain.

\subsubsection{Combination Label Construction}

Given the target domain label \begin{math} y^t \end{math} and the source domain aggregate label \begin{math} y^s \end{math}, the construction manner of the interest combination label \begin{math} y^{icl} \end{math} is shown in Table \ref{tab:ICL label}, where \begin{math} \eta \end{math} denotes the group consistency interest,  which will be discussed in the next subsection. Here we explained why the ICL is constructed in the manner of Table \ref{tab:ICL label}.
\begin{table}[h]
  \caption{Interest Combination Label}
    \vspace{-0.2cm}
  \label{tab:ICL label}
  \begin{tabular}{cc|c}
    \toprule
    $y^t$ & $y^s$  & $y^{icl}$  \\
    \midrule
    0 & 0 & 0 \\
    1 & 0 & 1 \\
    0 & 1 & $\eta$ \\
    1 & 1 & 2 \\
    \bottomrule
  \end{tabular}
  \vspace{-0.2cm}
\end{table}

\begin{itemize}
\item When \begin{math} y^t = 0, y^s = 0 \end{math} or \begin{math} y^t = 1, y^s = 0 \end{math}, there is no need to transfer interest from the source domain as the user has no interest in the source domain. Through the supervision of ICL, the interest transfer probability learned by the model will tend to 0 when training loss is minimized.
\item When \begin{math} y^t = 1, y^s = 1 \end{math}, it indicates that the user is likely to purchase the same item in both the source and target domains. In this case, the source domain interest is suitable for transferring to the target domain, which is a positive transfer situation. Thus the ICL is set to 2 to make the model learn a high interest transfer probability, ideally close to 1.
\item When \begin{math} y^t = 0, y^s = 1 \end{math}, it is unclear whether the user's source domain interest can be transferred to the target domain as user has different interests in the source and target domains. At this point, we extract the consistency of the user group's interest within the source and target domains as a supervised signal to determine whether the interest can be transferred across domains. Thus, the ICL is represented as \begin{math} y^{icl} = y^t + \eta \end{math}, that is, user's target domain interest \begin{math} y^t \end{math} plus group consistency interest \begin{math} \eta \end{math}. The interest transfer probability learned by the model will tend to \begin{math} \eta \end{math} by the ICL.
\end{itemize}

\subsubsection{Group Consistency Interest}
When a user's interests in the source and target domains are inconsistent, the group consistency interest is used to represent the interest that is suitable for transferring to the target domain. Consider the following scenario, many users have purchased the same item in both the source and target domains. Even if a user has not purchased the item in the target domain, it does not necessarily indicate a lack of interest in the item. This could be due to the interest estimation bias of the recommendation system, resulting in no exposure to the item for this user. In this case, the group consistency interest obtained by mining collaborative filtering information can more accurately reflect the transferability of cross-domain interests. The group consistency interest of item \begin{math} i \end{math}  can be calculated using the following formula:
\begin{equation}
    \eta_i= \frac{U_{pay_i}^t \cap U_{pay_i}^s}{U_{pay_i}^t \cup U_{pay_i}^s}
    \label{eq: group_interest}
\end{equation}
where \begin{math} U_{pay_i}^t  \end{math} and \begin{math} U_{pay_i}^s  \end{math} denote the set of users who have paid for item \begin{math} i \end{math} in the target domain and source domain, respectively. Notably, the computation of Equation~\ref{eq: group_interest} is time-efficient with the Spark tool\footnote{\url{https://spark.apache.org}} and only requires daily updates when deployed in real-world applications.


\subsection{Scene Selector Network}

Users' interest preferences in Meituan have a strong correlation with specific contexts, as most users prefer to purchase food delivery on weekdays and choose in-store services on weekends. Scene information plays a significant role in predicting user interests and is also an important factor affecting cross-domain interest transfer. Therefore, we propose SSN in the EXIT framework to learn the transfer intensity of different interests from the source domain to the target domain under fine-grained scenes, in order to ensure that the interest transfer probability matches the real changes in user interests. For instance, we expect a high interest transfer probability for food delivery in the scenario of <weekday, midday, office building, and white-collar>, and a relatively low transfer probability for in-store services in this scenario.

The SSN first concatenates the user embedding \begin{math} \mathbf{E}^U \end{math}, item embedding \begin{math} \mathbf{E}^M \end{math}, and context embedding \begin{math} \mathbf{E}^C \end{math}, and then uses a fully connected layer to compress the concatenated embedding to \begin{math} \mathbf{E}^{hid} \end{math}. SSN takes the compressed embedding \begin{math} \mathbf{E}^{hid} \end{math} and the scene embedding \begin{math} \mathbf{E}^{scene} \end{math} as inputs to enhance the utilization of scene information, where \begin{math} \mathbf{E}^{scene} \end{math} is concatenated by the embeddings of various scene features such as user's age, item type, current location, and current time.  Finally, SSN outputs the interest transfer probability \begin{math} P_{trans} \end{math}, which is formulated as follows:
\begin{equation}
    \mathbf{E}^{hid}= FC([\mathbf{E}^U || \mathbf{E}^M || \mathbf{E}^C])
\end{equation}
\begin{equation}
    \mathbf{E}^{scene}= [\mathbf{E}(F_{age}) || ... || \mathbf{E}(F_{type}) || ...  || \mathbf{E}(F_{hour})])
\end{equation}
\begin{equation}
    \mathbf{H}^{SSN}= MLP([\mathbf{E}^{hid} || \mathbf{E}^{scene}])
\end{equation}
\begin{equation}
    P_{trans}= Sigmoid(FC(\mathbf{H}^{SSN}))
\end{equation}
where \begin{math} || \end{math} denotes the vector concatenation operation. \begin{math} \mathbf{E}(\cdot) \end{math} denotes the embedding layer. \begin{math} F_{age}, F_{type} \end{math} and \begin{math} F_{hour}\end{math} represent three different scene features. \textit{FC} denotes the fully connected layer that performs the dimensional transformation of the dense embedding. \textit{MLP} refers to a set of DNN layers with activation functions, and \textit{Sigmoid} is the activation function with output values between [0,1].

\subsection{Multi-Interest Joint Loss}
The supervised labels for IPN modeling target domain interest and source domain interest are \begin{math} y^{t} \end{math} and \begin{math} y^{s} \end{math} respectively, which are true/false flags. For the classification problem, this paper employs the classical cross-entropy loss. The supervised signal for modeling the interest transfer intensity of SSN is provided by the ICL, and SSN is trained in coordination with the multi-task network. As shown in Table \ref{tab:ICL label},  ICL is a continuous value, so we use L1 loss to measure the distance between the predicted interest and ICL. Thus, the multi-interest joint loss is represented as follows:
\begin{equation}
    \begin{split}
    Loss_{joint}  &= {\lambda}_1 Loss_{target} + {\lambda}_2 Loss_{source} + {\lambda}_3 Loss_{icl} \\
    &={\lambda}_1 \sum_{x_i \in X^t}^{N_t} CrossEntropy\left(y_i^t, P_i^{t}\right) \\
    &+{\lambda}_2 \sum_{x_i \in X^t}^{N_t} CrossEntropy\left(y_i^{s}, P_i^{s} \right) \\
    &+{\lambda}_3 \sum_{x_i \in X^t}^{N_t} \left| y_i^{icl} -  \left(P_i^{t} + P_i^{s}*P_i^{trans} \right)\right| 
\end{split}
\end{equation}
where \begin{math} {\lambda}_1, {\lambda}_2 \end{math} and \begin{math} {\lambda}_3 \end{math} are weights that control the importance of different losses. \begin{math} x_i \end{math} is the training sample from training set \begin{math} X^t \end{math} and \begin{math} N_t \end{math} is the number of training set samples. \textit{CrossEntropy} denotes cross-entropy loss function. \begin{math} y_i^t, y_i^s \end{math} and \begin{math} y_i^{icl} \end{math} denote the target domain label, source domain aggregate label, and interest combination label, respectively. While \begin{math} P_{i}^t, P_{i}^s \end{math} respectively denote the user's target domain interest and source domain interest predicted by IPN, and \begin{math} P_i^{trans} \end{math} denotes the interest transfer probability output by SSN.

\subsection{Online Serving}
When in online serving, the EXIT outputs the user's complete interest, including target domain interest as well as interest transferred from the source domain. Since the output of EXIT has turned into the summation of two different parts, there is no guarantee of the output range. Hence, we truncate the final prediction between [0,1] with the below equation:
\begin{equation}
    P'_{whole} = min\{1, P_{whole}\}
    \label{eq:online serving}
\end{equation}

\section{Experiments}
\subsection{Datasets}

 \subsubsection{Large-Scale Industrial Offline Dataset}

 We collect the user exposure and purchase logs in the homepage recommendation, search, and channel section from the Meituan App, to build the real-world industrial dataset. The target domain is the homepage recommendation domain and the source domains are the search domain and channel section domain. The statistics of the dataset are shown in Table \ref{tab:offline_dataset}. Although datasets such as Amazon\footnote{\url{http://jmcauley.ucsd.edu/data/amazon/index_2014.html}} are publicly available, their source domain and target domain do not share users and items simultaneously. Hence, we select this in-house industrial data to evaluate the proposed model, which is larger than the public datasets available.

{\small
\begin{table}[t]
  \caption{Statistics of the industrial offline dataset. \begin{math} D_t \end{math}: target domain. \begin{math} D_s \end{math}: source domain.}
    \vspace{-0.2cm}
  \label{tab:offline_dataset}
    \begin{tabularx}{0.48\textwidth}{l@{\hspace{0.22cm}} l@{\hspace{0.2cm}} l@{\hspace{0.2cm}} l@{\hspace{0.2cm}} X X}
    \Xhline{1pt}
    Dataset &\#Samples & \#Users& \#Items & \#Purchases \newline (\begin{math}D_t \end{math})  & \#Purchases (\begin{math} D_s \end{math})  \\
    \midrule
    Train & 1,047,491,213 & 126,274,195 & 21,410,172 & 13,283,510 & 29,777,834 \\
    Test & 260,800,669 & 53,307,975 & 12,481,469 & 3,097,937 & 7,163,564 \\
    \Xhline{1pt}
  \end{tabularx}
    \vspace{-0.4cm}
\end{table}
}

 \subsubsection{Online Dataset}
The introduction of source domain knowledge causes the model to learn source domain interests that do not exist in the target domain, thus expanding the prediction interest space of the model and leading to the exposure of some items that could not be exposed before. For this reason, single-domain and cross-domain methods have different prediction interest spaces. Evaluating the model solely on the offline dataset in the target domain interest space does not fully reflect the advantages of the cross-domain methods. To ensure a more accurate evaluation of the performance of different methods, we deploy EXIT and the baseline methods in the Meituan homepage recommendation production environment, which contains hundreds of millions of user requests. The source and target domains correspond with the offline dataset. The users are randomly divided into different buckets of similar volume and are assigned our proposed model or baseline models for selecting items to be exposed in Meituan homepage. Each bucket involves about one million users. The size of the item candidate set is about 100 million, covering the most active products at Meituan.

\subsection{Compared Baselines}
To demonstrate the superiority of our EXIT framework, we compare it with various baselines, including single-domain and cross-domain methods. The single-domain baseline methods include LR~\cite{RichardsonDR07}, DNN~\cite{he2017neural}, DeepFM~\cite{GuoTYLH17}, and DCN~\cite{WangFFW17}. LR is a linear model, while DNN is a deep nonlinear model. DeepFM extracts both low-order and high-order features simultaneously, while DCN performs explicit feature crossing. The cross-domain baseline methods are MV-DNN~\cite{ElkahkySH15}, CoNet~\cite{HuZY18}, MiNet~\cite{Ouyang20}, STAR~\cite{ShengZ21} and UniCDR~\cite{cao2023towards}.

\subsection{Evaluation Metrics}
For offline experiments, we use the widely used Area Under the ROC Curve (AUC) and Logloss as evaluation metrics~\cite{silveira2019, shani2011, GuoTYLH17, Ouyang20}. For online A/B tests, we use three most important evaluation metrics in the Meituan platform: Click-Through Conversion Rate (CTCVR), Gross Transaction Value (GTV), and Negative Feedback Rate (NFR). CTCVR represents the conversion efficiency of the online system, calculated as the total payment volume on the platform divided by the total exposure volume. GTV represents the total amount of business conducted on the platform. NFR evaluates user satisfaction with recommendation results and is a key indicator of user experience. For CTCVR and GTV, the larger the better. For NFR, the smaller the better.

\subsection{Parameter Settings}
\begin{figure}[t]
\setlength{\abovecaptionskip}{-1pt} 
\setlength{\belowcaptionskip}{-1pt}
\centering
\includegraphics[width=0.45\textwidth]{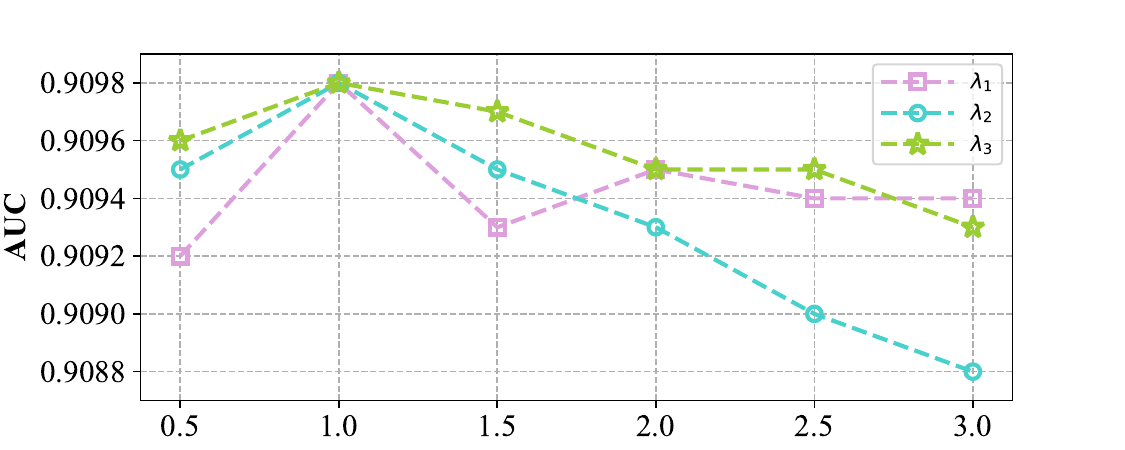} 
\caption{Parameter sensitivity of \begin{math} \lambda_1,\lambda_2\end{math} and  \begin{math} \lambda_3 \end{math}.}
\label{fig:sensitivity}
\vspace{-0.4cm}
\end{figure}

We set the dimension of the embedding vectors for each feature as embedding dim = 8. The batch size is set to 384. All the methods are implemented in TensorFlow and optimized by the Adam algorithm~\cite{KingmaB14}. Each method runs 3 times in the offline dataset and each online A/B test runs for a period of 7 days, with the average results being reported. For our proposed EXIT, we use MMoE~\cite{MaZYCHC18} with two shared experts as the multi-task network. The target tower and source tower are both 2-layer DNNs with PReLU activation functions. In SSN, we utilize the following scene features: user-side scene features such as gender, occupation, and age; item-side features such as item level 1 category, item level 2 category, item level 3 category, and item business; and context-side scene features such as hour, day of the week, request page, and connection type. We conducted parameter sensitivity experiments on \begin{math} \lambda_1,\lambda_2\end{math} and  \begin{math} \lambda_3 \end{math} in the multi-interest joint loss. The results are showin in Figure~~\ref{fig:sensitivity}. We found it easy to adjust the parameters to appropriate values. In practice, we set \begin{math} \lambda_1,\lambda_2\end{math} and \begin{math} \lambda_3 \end{math} all to 1.

\begin{table}[t]
  \caption{Offline and online experimental results on different methods. "\textuparrow": the larger the better. "\textdownarrow": the smaller the better. \underline{Underline}: runner-up.}
  \vspace{-0.2cm}
  \label{tab:experimental_results}
  \begin{tabular}{c@{\hspace{0.20cm}} c@{\hspace{0.18cm}} c@{\hspace{0.20cm}}| c@{\hspace{0.19cm}} c@{\hspace{0.19cm}} c@{\hspace{0.19cm}}}
    \Xhline{1pt}
         \multirow{2}{*}{Method}&  \multicolumn{2}{c|}{Offline Metric} &  \multicolumn{3}{c}{Online Metric}\\ \cline{2-6}
 & AUC\textuparrow & Logloss\textdownarrow & CTCVR\textuparrow & GTV\textuparrow &NFR\textdownarrow \\ 
    \hline
         LR~\cite{RichardsonDR07}& 0.8984 & 0.06861 & - & - & - \\ 
         DNN~\cite{he2017neural}& 0.9052 & 0.06677 & - & - & - \\  
         DeepFM~\cite{GuoTYLH17}& 0.9069 & 0.06685 & +0.278\% & +1.374\% &  \underline{-1.245\%} \\ 
         DCN~\cite{WangFFW17}& 0.9067 & 0.06675 & +0.448\% & +0.896\% & -0.921\% \\ \hline 
         MV-DNN~\cite{ElkahkySH15} & 0.9008 & 0.06820 & -0.024\% & +0.858\% & +2.025\% \\ 
CoNet~\cite{HuZY18} & 0.9073 & 0.06662 & +0.540\% & +1.616\% & +0.272\% \\
 MiNet~\cite{Ouyang20} & \underline{0.9078} &  \underline{0.06655} & +0.650\% & +2.174\% & -0.251\% \\ 
 STAR~\cite{ShengZ21} & 0.9069 & 0.06672 & +0.520\% & +1.199\%& -0.781\% \\
 UniCDR~\cite{cao2023towards} & 0.9075 & 0.06668 & \underline{+0.775\%} & \underline{+2.718\%} & -1.211\% \\ \cline{1-6}
 \textbf{EXIT(ours)}& \textbf{0.9098} & \textbf{0.06637} & \textbf{+1.336\%} & \textbf{+4.109\%} & \textbf{-6.937\%}\\ 
  \Xhline{1pt}
  \end{tabular}
    \vspace{-0.3cm}
\end{table}

\subsection{Overall Performance Comparison}
We evaluate the performance of EXIT and baseline methods on both the offline dataset and online A/B tests. For the online experiments, we set DNN as the benchmark and report the performance improvements of other methods compared to DNN due to business privacy. As illustrated in Table \ref{tab:experimental_results}, the consistent improvement validates the effectiveness of EXIT. It is observed that deep models have significant performance improvements than LR. DeepFM and DCN outperform DNN, indicating that feature combination and feature crossing can improve prediction performance. As to the cross-domain baseline methods, they show improvements over single-domain methods on AUC and Logloss, demonstrating the benefits of knowledge transfer. In terms of the online CTCVR and GTV, cross-domain methods achieve more significant improvements. This suggests that evaluating offline performance only in the target domain interest space cannot fully reflect the actual performance of CDR methods, verifying the necessity of conducting online comparison experiments. However, for the NFR metric that measures user experience, most cross-domain baselines perform worse than DeepFM and DCN. This is because these cross-domain methods do not explicitly consider the negative transfer issue, which results in the introduction of inappropriate interests into the target domain, consequently harming the user experience. Our proposed EXIT, by explicitly modeling effective interest signals that are suitable for transferring to the target domain, not only captures user interests more comprehensively but also prevents negative transfer. Therefore, EXIT significantly outperforms both the single-domain and cross-domain baseline methods on all offline and online evaluation metrics.

\begin{table}
\centering
  \caption{The ablation study for the components in EXIT. The online results are performance comparisons with DNN.}
  \label{tab:ablation_study}
    \vspace{-0.2cm}
  \begin{tabular}{c@{\hspace{0.20cm}} c@{\hspace{0.18cm}} c@{\hspace{0.20cm}}| c@{\hspace{0.19cm}} c@{\hspace{0.19cm}} c@{\hspace{0.19cm}}}
  \Xhline{1pt}
       \multirow{2}{*}{Ablations}&  \multicolumn{2}{c|}{Offline Metrics}&  \multicolumn{3}{c}{Online Metrics}\\ \cline{2-6}
 & AUC\textuparrow & Logloss\textdownarrow & CTCVR\textuparrow & GTV\textuparrow & NFR\textdownarrow \\  
    \hline
    w/o ICL & 0.9055 & 0.06749 & +0.581\% & +1.226\% & +0.779\% \\
    w/o SSN & 0.9075 & 0.06661 & +1.092\% & +1.971\% & -4.159\% \\
    w/o joint loss & \multicolumn{2}{c|}{not converged} & - & - & - \\
        \hline
    ICL$(\eta=0)$ & 0.9081 & 0.06643 &+0.874\% & +3.053\% &-5.231\% \\  
    ICL$(y^t+\eta)$ & 0.9074 & 0.06658 & +0.544\% & +1.814\% & -3.670\% \\  
       \hline
    \textbf{EXIT(ours)} & \textbf{0.9098} & \textbf{0.06637} & \textbf{+1.336\%} & \textbf{+4.109\%} & \textbf{-6.937\%}\\ 
  \Xhline{1pt}
  \end{tabular}
      \vspace{-0.2cm}
\end{table}


\renewcommand{\arraystretch}{1.1} 
\begin{table*}[t]
  \caption{Case study of an actual Meituan user. \begin{math} D_{t} \end{math} represents the target domain and \begin{math} D_{s} \end{math} represents the source domain. Y/N respectively indicate if the item has been exposed or not exposed to the user.}
  \label{tab:case_study}
        \vspace{-0.3cm}
  \begin{tabular}{p{3cm} p{3cm} c c c c c c c}  
    \Xhline{1pt}
         \multirow{2}{3cm}{Purchase behaviors in \begin{math} D_{t} \end{math}}&  \multirow{2}{3cm}{Purchase behaviors in \begin{math} D_{s} \end{math}}&  \multirow{2}{*}{Candidate Items/ Category}&  \multicolumn{4}{c}{Model Scores} & \multirow{2}{*}{Exposure}\\ \cline{4-7}
 & & &  \begin{math} P_{target} \end{math} & \begin{math} P_{source} \end{math} & \begin{math} P_{trans} \end{math} & \begin{math} P_{whole} \end{math} \\  
    \midrule
         \multirow{4}{3cm}{Food delivery 3 times; Hamburger deal once}    &  \multirow{4}{3cm}{Hot pot deal 4 times; Barbecue deal 3 times; Medicine 2 times}  & Item1 / Hot pot deal & 0.047 & 0.798 & 0.792  & 0.679 & Y \\  \cline{3-8}
         &   & Item2 / Barbecue deal & 0.010 & 0.759 & 0.614 & 0.476 & Y  \\  \cline{3-8}
         &   & Item3 / Cold medicine & 0.076 & 0.562 & 0.048 & 0.103 & N  \\   \cline{3-8}
         &   & Item4 / Cold medicine & 0.064 & 0.653 & 0.011 & 0.071 & N  \\   
    \Xhline{1pt}
  \end{tabular}
  \vspace{-0.3cm}
\end{table*}

\subsection{Ablation Study}
We conducted several ablation studies and the ablation experimental results in Table \ref{tab:ablation_study} show that every component in EXIT is essential.

 \subsubsection{Effect of the Main Components.}
Firstly, we eliminated the ICL and indiscriminately incorporated all interest signals from the source domain into the target domain. As shown in Table \ref{tab:ablation_study}, the performance of both offline and online experiments exhibits a significant decline. Notably, the NFR metric shows a substantial increase, which may be attributed to the introduction of interest signals that are inappropriate for the recommendation domain. The results suggest that the ICL not only assists in more accurately modeling user interests but also plays a crucial role in preventing negative transfer. Secondly, to figure out the effectiveness of SSN, we eliminate the additional scene features and scene embeddings. The experimental results convey that scene information is beneficial for modeling the interest transfer intensity. Moreover, to investigate the function of the multi-interest joint loss, we removed the cross-entropy losses associated with \begin{math} y^{t},y^{s} \end{math} and eliminated the stop gradient from the network in Figure~\ref{fig:framework}, leaving only the L1 loss associated with the final ICL to train the model. However, the training process fails to converge without the multi-interest joint loss. This indicates that learning the source domain interest and target domain interest separately is the fundamental task of the model, which also aligns with the global perspective in Equation \ref{eq: global perspective}.

 \subsubsection{Effect of the Group Consistency Interest.}

We study the role of group consistency interest through two variants of ICL: (1) \begin{math} ICL(\eta=0) \end{math} indicates that ICL is constructed solely based on user's personalized interests, without considering group consistency interests; (2) \begin{math} ICL(y^t+\eta) \end{math} indicates the use of group consistency interests to represent the transferability of cross-domain interests in all scenarios. The experimental results in Table~\ref{tab:ablation_study} demonstrate that ignoring group consistency interests or relying solely on them both result in a decline in model performance. Additionally, user's personalized interests have a more significant impact on ICL. The results confirm the soundness of the ICL construction method. To accurately represent the transferability of cross-domain interests, we propose utilizing group consistency interests when the user's source and target domain interests are inconsistent. When interests align in both domains, the decision to transfer interests should be based on the user's personalized interests.

\subsection{Online Deployment}
We deployed EXIT on the homepage recommendation system of Meituan App and conducted online A/B tests over two weeks during Mar.2024. Since the base production model is a multi-task model that simultaneously models multiple objectives, we maintain the structure of all other objectives and only replace the modeling of the CTCVR target with the EXIT framework during online deployment. The introduction of EXIT brings 1.23\% overall CTCVR lift, 3.65\% GTV lift, and 6.62\% NFR decrease. The results demonstrate the effectiveness of the EXIT in the actual production recommendation system. Currently, EXIT has been deployed for handling major online traffic at the homepage on Meituan App, which is one of the top mobile apps in our country. 

\section{Case Study}

To further verify the working mechanism of our explicit framework, we provide case studies from the online recommendation system. Table \ref{tab:case_study} shows the recent behaviors of an actual Meituan user in the Meituan App and how the EXIT scores different candidate items during an online request from this user. The user exhibits diverse interests across different business domains, primarily purchasing food delivery in the target domain and purchasing group deals and medicines in the source domain. Therefore, it is difficult to accurately model user interests with only single-domain data. Furthermore, we examine how EXIT scores candidate items during the online request. For group deals like hot pot or barbecue that the user is interested in within the source domain, both \begin{math} P_{source} \end{math} and \begin{math} P_{trans} \end{math} are found to be quite high. According to Equation~\ref{eq: global perspective}, higher values of both \begin{math} P_{source} \end{math} and \begin{math} P_{trans} \end{math} result in higher predicted CTCVR scores. Consequently, the model prioritizes and exposes these items to the user. This suggests that EXIT is capable of effectively transferring knowledge across domains, thus facilitating a more accurate understanding of user interests. As for the candidate items of cold medicine, although the user may have previously purchased cold medicine in the source domain due to sickness or other reasons, the group consistency interest for such items is minimal. By supervising the model learning with the ICL, it learns a lower \begin{math} P_{trans} \end{math}, thereby omitting such source domain instant interests during interest transfer. By controlling the interest transfer probability, EXIT can effectively identify interest signals that are inappropriate for the target domain, thereby preventing negative transfer.

\section{Conclusion}
In this paper, we propose a simple and effective EXplicit Interest Transfer framework called EXIT, to tackle the challenges in CDR. EXIT constructs Interest Combination Label (ICL) for supervised learning of the cross-domain interest transfer, which provides a novel explicit paradigm for CDR and alleviates the issue of negative transfer. To collaborate with the ICL, we propose a scene selector network to model the intensity of interest transfer in fine-grained scenes. By this means, EXIT can accurately and completely capture user interests and improve the recommendation performance. Extensive offline and online experiments demonstrate the superiority of EXIT. EXIT has been successfully deployed to the Meituan homepage recommendation system, bringing significant business benefits to Meituan. We anticipate that EXIT will inspire more efforts in the future towards explicit cross-domain recommendations.  In the future, we will explore how supervised learning can be used to model knowledge transfer in domains without item or user overlap,  to achieve broader applications of the explicit paradigm. 
\bibliographystyle{ACM-Reference-Format}
\balance
\bibliography{sample-base}

\end{document}